\documentclass{camera}
\usepackage{graphicx}

\def\la{\mathrel{\mathchoice {\vcenter{\offinterlineskip\halign{\hfil
$\displaystyle##$\hfil\cr<\cr\sim\cr}}}
{\vcenter{\offinterlineskip\halign{\hfil$\textstyle##$\hfil\cr<\cr\sim\cr}}}
{\vcenter{\offinterlineskip\halign{\hfil$\scriptstyle##$\hfil\cr<\cr\sim\cr}}}
{\vcenter{\offinterlineskip\halign{\hfil$\scriptscriptstyle##$\hfil\cr<\cr
\sim\cr}}}}}

\newcommand{\etal} {{\it et al.}}
\newcommand{\eas} {{\sc eas }}
\newcommand{\cray} {{\sc cr }}
\newcommand{\crays} {{\sc cr}s }

\begin{document}
    
\title{THE CHEMICAL COMPOSITION OF COSMIC RAYS \footnote{Invited 
paper presented at Vulcano Workshop 2002: {\sl ``Frontier Objects in 
Astrophysics and Particle Physics'',\newline
e-mail: Karl-Heinz.Kampert@ik.fzk.de}}}

\author{Karl-Heinz Kampert}

%
\organization{
Institut f\"ur Experimentelle Kernphysik, University of Karlsruhe,\\
and Institut\ f\"ur Kernphysik, Forschungszentrum Karlsruhe,\\
P.O. Box 3640, 76021~Karlsruhe, Germany}

\maketitle

\abstract{A brief review about the chemical composition of cosmic
rays in the energy range $10^{15} \la E \la 10^{20}$~eV is given. 
While there is convincing evidence for an increasingly heavier
composition above the knee, no clear picture has emerged at the
highest energies, yet.  We discuss implications about the origin
of cosmic rays and emphasize systematic differences related to
data analysis techniques and to the limited understanding of the
air shower development.}

\section{Introduction}

When being asked to formulate the most crucial, yet unanswered
questions about {\sc cr}s, one likely would end up with a list
as such:
\begin{enumerate}
    \item Where and what are the sources of high energy {\sc
    cr}s?
    \item How does the transport from the source(s) to the
    solar system affect their properties?
    \item What is the origin of structures observed in their
    energy distributions?
    \item Is there an upper end to the energy spectrum?

\end{enumerate}
The enormous interest particularly in the highest energy {\sc
cr}s originates also from the fact that they provide a direct
link to the problems of understanding the evolution of the early
universe and to physics beyond the standard model of particle
physics in an energy range not accessible to man-made
accelerators.

The chemical composition of \crays measured as a function of
their energy provides a key to the answers of above questions and
large efforts have been undertaken to provide reliable
experimental data for comparison with models of \cray origin.  In
fact, direct \cray measurements performed at energies up to
energies of several GeV have provided a great deal of information
about the source composition, propagation effects, the `age' of
{\sc cr}s, etc.  However, for \eas experiments the situation has
been proven much harder than expected.  This is mostly because of
the large fluctuations of \eas parameters and because of
uncertainties in corresponding \eas simulations.  Nevertheless,
significant progress has been made in recent years and we have
reached a point where high quality \eas data {\em and} models of
\cray origin and acceleration can be confronted to each other. 
The information extracted from such measurements may be
complemented by high energy (TeV) gamma and neutrino radiation. 
The latter messengers point directly to the source but are only
secondary probes since they originate from interactions of \crays
with the local environment at the source.  Understanding the
common picture of photons, neutrinos and high quality \cray
energy and composition data may finally yield a conclusive
picture to the open {\sc cr}-questions of the century.

\section{Experimental Results}

The energy spectrum of \crays follows a simple power law
behaviour ($dJ/dE \propto E^{-\gamma}$) over many orders of
magnitude.  There are only two well established structures in the
\cray spectrum; the {\it knee} at $E \simeq 3$ PeV where the
index changes from $\gamma \simeq 2.75$ to 3.1 and the {\it
ankle} at $E \simeq 5$ EeV where the spectrum turns up again. 
The origin of the structures, however, remains unknown as is the
existence or non-existence of the Greisen-Zatsepin-Kuzmin
cut-off.  Since structures in a spectrum generally contain
important information, many experiments have focussed to study
these energy ranges.  However, due to the low flux of high energy
{\sc cr}s, all of these structures are accessible only to \eas
experiments.  The primary energy and composition is then inferred
either from particle measurements at ground or from Cherenkov and
fluorescence light observations, or from combinations of both. 
Major particle arrays in operation or still delivering newly
analysed data include:
{\sc agasa}, 
{\sc casa-mia}, 
{\sc eas-top}, 
{\sc grapes}, 
Haverah Park, 
{\sc hegra}, 
{\sc kascade}, 
{\sc Maket-Ani}, 
{\sc msu}, 
{\sc spase}, and 
Tibet-As$\gamma$. 
{\sc Eas} experiments exploiting optical observations include:
{\sc blanca}, 
{\sc dice}, 
Fly's Eye, 
{\sc hegra}, 
{\sc HiRes}, 
and {\sc tunka}. 
A brief overview about most experiments and their observational
techniques can be found in Refs.~\cite{kampert01,swordy02}.

Previous results about the \cray composition were reported e.g.\
in \cite{castellina01,hoerandel02}.  The traditional and perhaps most
sensitive technique to infer the \cray composition from \eas data
is based on measurements of the electron ($N_e$) and muon numbers
($N_\mu$) at ground.  For consistency tests or for improving the
sensitivity, $N_{e}$ and $N_{\mu}$ may be complemented by
additional \eas parameters, such as the electron lateral
distribution ({\sc ldf}), muon production height, hadronic
observables, particle arrival times, etc.

Cherenkov and fluorescence observations are restricted to clear
moonless night which imposes a maximum duty cycle of approx.\
10\,\%.  In non-imaging Cherenkov measurements the composition is
inferred from the steepness of the lateral Cherenkov light
intensity at distances up to 120~m from the shower core.  Instead
of presenting the Cherenkov {\sc ldf} for comparison with {\sc
eas} simulations, many experiments have preferred to introduce
another intermediate step by inferring the maximum of the
longitudinal shower development, $X_{max}$, by means of \eas
simulations so that only in a second step these inferred
$X_{max}$ values are compared with \eas simulations.  Obviously,
this introduces additional systematic uncertainties and may
result in inconsistencies when for example different hadronic
interaction models are applied to the (Ch-light intensity
$\leftrightarrow$ energy )- and (Ch {\sc ldf} $\leftrightarrow$
$X_{max}$)-relations on the one hand and to the $X_{max}$
distributions of \eas simulations employing different hadronic
interaction models on the other.  Up to now, only fluorescence
measurements allow direct measurements of $X_{max}$ from
observations of the longitudinal shower profile.

\begin{figure}[t]
\centerline{\includegraphics[width=10.0cm]{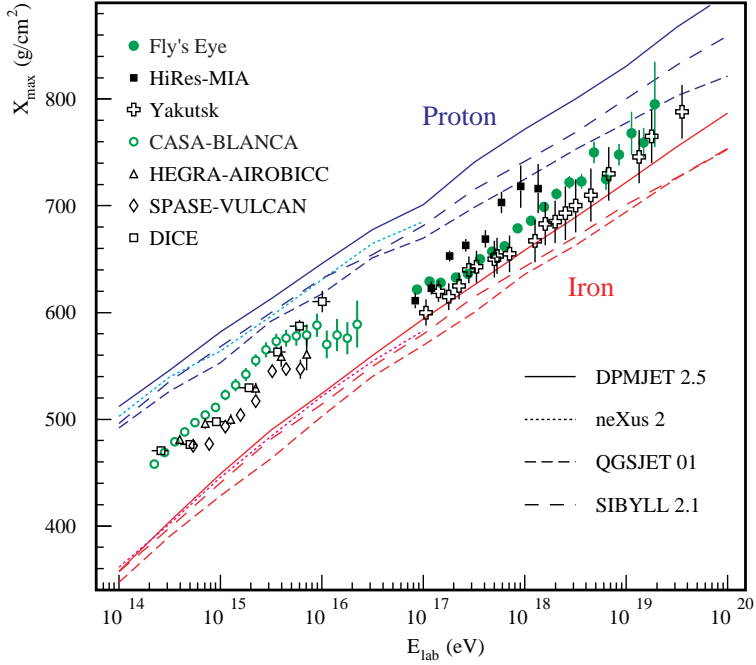}}
\caption[]{Compilation of $X_{max}$ from different experiments 
compared with {\sc corsika} simulations using different hadronic 
interaction models (adopted from Ref.~\cite{heck01}).}
\end{figure}

\begin{figure}[t]
\centerline{\includegraphics[width=12cm]{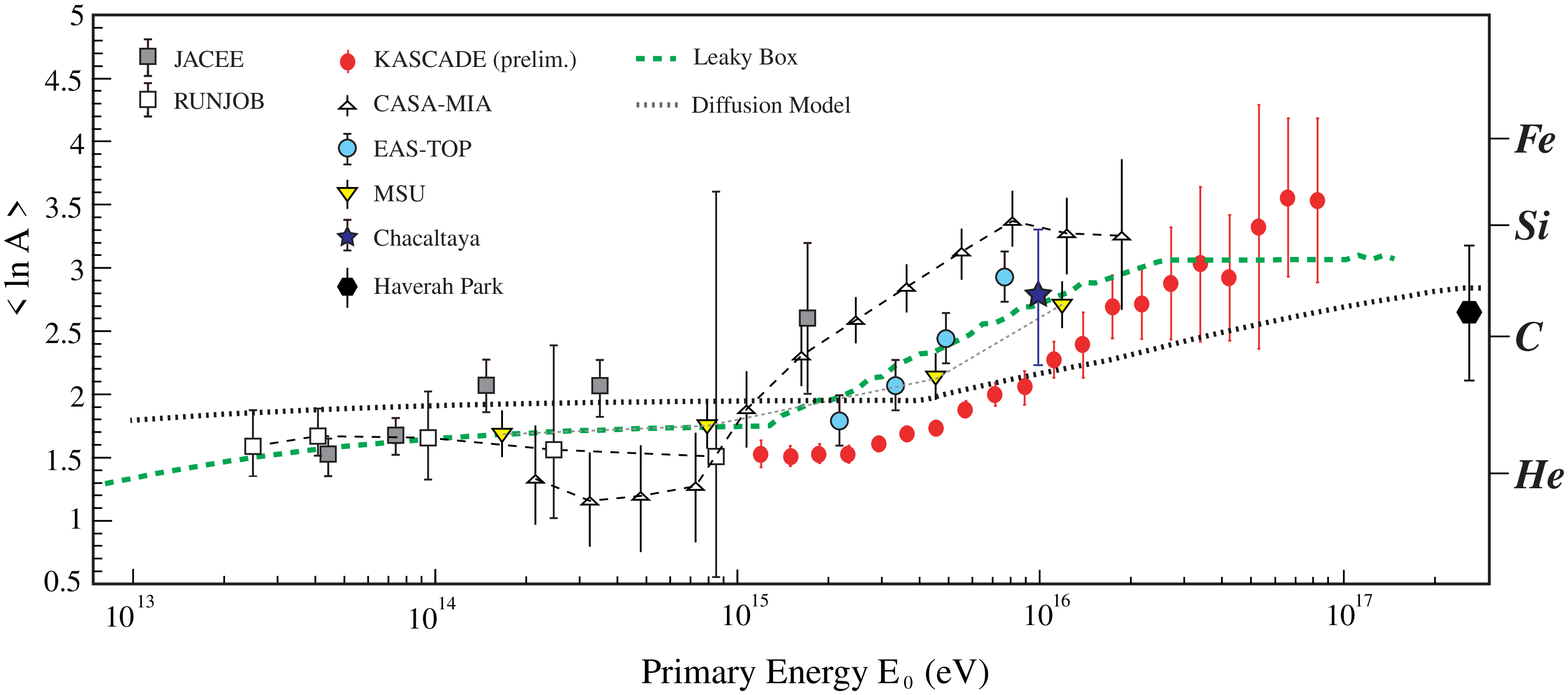}}
\caption[]{Mean logarithmic mass from direct \cray measurements
and air shower arrays.  Data are from {\sc Jacee}
\cite{asakamori98}, {\sc Runjob} \cite{runjob-01}, {\sc kascade}
\cite{ulrich-02}, {\sc casa-mia} \cite{glasmacher99}, {\sc
eas-top} \cite{alessandro01}, the {\sc msu} array \cite{fomin96},
Chacaltaya \cite{aguirre00}, Haverah Park \cite{ave02} and are
compared to a simplified leaky box \cite{swordy-95} and diffusion
model \cite{maurin-02}.}
\end{figure}

A compilation of experimental results on $X_{max}$ is shown in
Fig.~1 for a wide range of primary energies \cite{heck01}.  The
data are compared to {\sc corsika} simulations using different
hadronic interaction models.  As can be seen from that figure,
differences between the models are on the order of 25~g/cm$^2$ in
the knee region and increase to about 40~g/cm$^2$ at higher
energies.  Systematic uncertainties in the data are estimated to
at least be of similar size.  Again, besides the shifts seen in
the simulated $X_{max}$-lines, also the experimental
$X_{max}$-values of the Cherenkov data may shift when extracted
using different interaction models.  With these uncertainties
kept in mind, all data seem to suggest a mixed composition over
the whole range of energies with no clear picture emerging about
a change in the composition.  Only the {\sc casa-blanca}
\cite{fowler01} and {\sc spase-vulcan} \cite{dickinson99} data
may indicate a lighter composition towards the knee with a
turnover to a heavier one above the knee.  The Cherenkov imaging
experiment {\sc dice} \cite{kieda99} appears to contradict that
picture.  However, electronic saturation effects were recently
discussed to have biased their data \cite{larsen01}.  The weak
indication from light to heavy when crossing the knee is followed
by a trend back towards a mixed-light composition in the energy
range $10^{17}$ - $10^{18}$ eV in the {\sc HiRes-Mia} data
\cite{abu-zayyad00}.  The original Fly's Eye data suggest a heavy
dominated composition with a much weaker change to a lighter
composition only around $E \simeq 10^{19}$ eV \cite{bird93}.  In
addition to experimental effects large uncertainties are added by
the poorly understood hadronic interaction models at these high
energies.

Data from \eas arrays are mostly presented in terms of the mean
logarithmic mass, $\langle \ln A \rangle = \sum_{i} r_{i} \ln
A_{i}$.  Despite the drawbacks of this observable
\cite{kampert01} we shall follow this approach to ease the
comparison of data from different experiments.  Figure~2 shows a
compilation of several ground arrays.  Data from the balloon
borne experiments {\sc Jacee} \cite{asakamori98} and {\sc Runjob}
\cite{runjob-01} and results from a simplified leaky-box
\cite{swordy-95} and diffusion model calculation assuming a knee
at a constant rigidity of 4 PV \cite{maurin-02} are included for
comparison.  All \eas arrays observe a change towards a heavier
composition across the knee.  However, absolute differences
between experiments up to $\langle \ln A \rangle \simeq 1$ are
found in the transition region.  As comparisons between {\sc
qgsjet} and {\sc neXus} indicate \cite{vanburen}, another overall
shift by $\langle \ln A \rangle \simeq 0.5$ should be added from
uncertainties in the interaction models.  The \eas data presented
in Fig.~2 were all analysed with reference to {\sc
corsika/qgsjet}.  Reasonably good agreement is seen also between
\eas data and direct measurements from the {\sc Runjob}
Collaboration favouring an almost constant composition from
approx.\ $10^{13}$ to $10^{15}$ eV. {\sc Jacee} on the other hand
suggests an increasingly heavier composition starting already at
approx.\ $10^{14}$~eV. The cause of this difference is not fully
understood.  Helium fluxes in {\sc Runjob} data are lower by a
factor of 2 as compared to {\sc Jacee} and, perhaps more
importantly, {\sc Jacee} attributes sub-Fe elements to the
Fe-group thereby increasing the $\langle \ln A \rangle$ values. 
The {\sc kascade} data shown in Fig.~2 result from an unfolding
procedure of the $N_{e}$- and $N_{\mu}^{tr}$-size distributions
\cite{ulrich-icrc01} and perhaps provide the best clue to the
knee puzzle.  The preliminary data convincingly show knee
structures in energy spectra selected by primary mass groups. 
Furthermore, their knee position changes towards higher energies
with increasing mass of the primaries.  Detailed examination of
the individual knee positions suggest a constant rigidity for all
components \cite{ulrich-icrc01,kampert-02} as is expected in
astrophysical acceleration and transport models.

The only composition analysis above $10^{17}$ eV based on a
ground array and using state of the art \eas simulations has
recently been presented by the Haverah Park group \cite{ave02}. 
Their analysis is based on the shape of the observed lateral
distribution function derived from the water-Cherenkov signals. 
Assuming a constant bi-modal composition in the energy range
0.2-1.0 EeV best agreement to the data is found for $(34 \pm
2)$\% protons and the rest iron.  Systematic uncertainties
introduced by the choice of hadronic models were estimated to
approx.\ 14\,\%.  This error is included in the data point shown
in Fig.~2.  Within error bars, this finding is in agreement to
the {\sc kascade} results at lower energies.  Furthermore, there
may a turnover back towards a medium-light composition at higher
energies.  Such a change is indeed reported by fluorescence
measurements of the {\sc HiRes} prototype working jointly with the {\sc
mia} muon array for energy determination \cite{abu-zayyad00} and
has been reported to some extend also by the Fly's Eye
collaboration \cite{bird93} (see above).  However, quantitatively
there are still significant differences in the composition at
energies above $10^{17}$~eV so that the question appears far from
being resolved.

\section{Final Remarks and Outlook}

A variety of experimental techniques is applied to infer the
energy and mass of \crays from {\sc eas}.  In the knee region,
two approaches have been proven successful: extended particle
arrays measuring electrons, muons, and high energy hadrons on the
one side and non-imaging Cherenkov measurements on the other.  At
the highest energies, the Cherenkov technique becomes impractical
but fluorescence observations take over.  The wide scattering of
reconstructed data points in any of the approaches, even when
employing the same hadronic interaction model, may hint to hidden
experimental systematics and/or to incomplete data analysis
techniques in at least some cases.  Indeed, the data analysis is
complicated by many subtle details: the energy spectrum is very
steep but (i) shower fluctuations are in general very large, (ii)
they depend strongly on the (to be reconstructed) mass of the
primary particle and, if not corrected for, (iii) their effect
mimics a primary dependent increase of fluxes which in addition
depends on the steepness of the underlying elemental spectra.  As
a consequence of the latter, the mass dependent changes of the
spectral indices at the individual knee positions causes a bias
towards a lighter composition above the knee.  Whether that
effect is responsible for the apparent dip towards light masses
seen in knee region of the Cherenkov data (see Fig.~1) cannot be
said at present.  Furthermore, on top of these complications,
experimental reconstruction accuracies generally dependent on the
primary energy and mass and have to be carefully corrected for as
well.  Clearly, sophisticated data analysis techniques are
required, such as unfolding procedures which do not make any
assumption about the shape of the individual primary energy
spectra or about their relative abundances.  Very promising
results have recently been presented by the {\sc kascade}
collaboration \cite{ulrich-icrc01,kampert-02} and the techniques
may be applied also to other existing experimental data.  The
present limitation of such approaches is, besides the available
Monte Carlo statistics, mostly given by the reliability of the
simulated tails in the used \eas observables.

The situation at higher energies becomes more complicated mostly
because of increasing uncertainties of interaction models. 
Akeno, Fly's-Eye \cite{dawson-98}, and {\sc HiRes-Mia} measure a
composition rich in Fe at $10^{17}$~eV which may become lighter
at higher energies.  On the other hand recent analyses of Haverah
Park data indicate a mostly light composition already above $2
\cdot 10^{17}$ eV \cite{ave02}. 

With these limitations in mind, what can be learned from the
present \eas data?  There appears convincing evidence for a
change from a `standard' light (p+He dominated) composition to a
heavy one (Si or heavier) in the energy range from the knee to
about $10^{17}$~eV. According to preliminary {\sc kascade} data,
this change in composition is caused by subsequent `breaks' in
the energy spectra of the different elements.  The data also
provide support for the knee being an effect of constant rigidity
at $E/Z \simeq 4\cdot10^{15}$~eV, such as is expected in
astrophysical models for confinement and/or acceleration in
magnetic fields.  The iron knee would then be expected at $E
\approx 10^{17}$~eV. It will be interesting to confront such data
with detailed model predictions.  If the knee in the spectra is
mostly due to galactic modulation, the question arises about to
which energies the Fe-group could dominate the all-particle flux
and to where the extragalactic component would take over.  As
recently argued by Berezinsky \etal, this transition could occur
at energies even lower than $10^{18}$~eV so that the composition
naturally would become light again.

To conclude, enormous progress has been made in recent years in
measuring the \cray composition more accurately than in earlier
generations of \eas experiments.  This is particularly true for
the energy range of the knee but new analyses have also been
reported at higher energies.  Besides further improving the data
analysis techniques, a wealth of new data at high energies will
become available in the very near future from {\sc
kascade-grande}, {\sc HiRes}, the Pierre Auger Observatory, and
in the more distant future possibly from space experiments like
{\sc euso} and {\sc owl} providing the required input for solving
the open questions about \crays.

\vspace*{5mm} {\bf Acknowledgement:} It is a pleasure to thank
the organizers of the Vulcano Workshop 2002 for their invitation
and for setting up such an interesting and fruitful meeting in a
very pleasent atmosphere.  This work has been supported in part
by the German Ministry for Research and Education.


\begin{thebibliography}{99}

\bibitem{kampert01}
K.-H.~Kampert, J. Phys. {\bf G27} (2001) 1663.

\bibitem{swordy02}
S.~Swordy \etal, Astropart. Phys. {\bf 18} (2002) 129.

\bibitem{castellina01}
A.~Castellina, Nucl. Phys. B (Proc. Suppl.) {\bf 97} (2001) 35.

\bibitem{hoerandel02}
J.~H\"orandel, Astropart. Phys. in press (astro-ph/0210453).

\bibitem{heck01}
D.~Heck \etal, 27$^{th}$ ICRC Hamburg, 1999, p233.

\bibitem{fowler01}
J.W.~Fowler \etal, Astropart. Phys. {\bf 15} (2001) 49.

\bibitem{dickinson99}
J.E. Dickinson \etal, {26$^{th}$ ICRC} Salt Lake City (1999) Vol. 3. p.136.

\bibitem{kieda99}
D.~Kieda \etal,
{26$^{th}$ ICRC Salt Lake City, Vol.\  {\bf 3}, p.\ 191}, 1999.

\bibitem{larsen01}
C.G.~Larsen, D.B.~Kieda, and S.P.~Swordy, 27$^{th}$ ICRC Hamburg, 
1999, p134.

\bibitem{abu-zayyad00}
T. Abu-Zayyad \etal, Phys. Rev. Lett. {\bf 84} (2000) 4276.

\bibitem{bird93}
D.J. Bird \etal, Phys. Rev. Lett. {\bf 71} (1993) 3401.

\bibitem{asakamori98}
K.~Asakamori \etal, Ap. J. {\bf 502} (1998) 278.

\bibitem{runjob-01}
A.V.~Apanasenko \etal,
Astropart. Phys. {\bf 16} (2001) 13

\bibitem{swordy-95}
S.~Swordy, {24$^{th}$ ICRC}, Rome, {\bf 2} (1995) 697

\bibitem{maurin-02}
D. Maurin, M. Casse, E. Vangioni-Flam, Astropart. Phys. (2002) in 
press.

\bibitem{vanburen}
J. van Buren, Diploma Thesis, University Karlsruhe (2002).

\bibitem{ulrich-02}
H.~Ulrich \etal, Nucl. Phys. B (Proc. Suppl.) 2002.

\bibitem{glasmacher99}
M.A.K.~Glasmacher \etal,
Astropart. Phys. {\bf 12} (1999) 1.

\bibitem{alessandro01}
B.~Alessandro \etal, {27$^{th}$ ICRC}, Hamburg, (2001) 124.

\bibitem{fomin96}
Yu.A.~Fomin \etal,
J. Phys. G {\bf 22} (1996) 1839.

\bibitem{aguirre00}
C.~Aguirre \etal,
Phys. Rev. D {\bf 62} (2000) 032003.

\bibitem{ave02}
M. Ave \etal, Astropart. Phys. in press; astro-ph/0203150.

\bibitem{ulrich-icrc01}
H.~Ulrich \etal, {27$^{th}$ ICRC}, Hamburg, (2001) 97

\bibitem{kampert-02}
K.-H. Kampert \etal,
{\it Invited, Rapporteur, and Highlight papers of ICRC 2001}, p240

\bibitem{dawson-98}
B.R. Dawson, R. Meyhandan, and K.M. Simpson,
Astropart. Phys. {\bf 9} (1998) 311,

\bibitem{berezinsky-02}
V. Berezinsky, A. Gazizov, and S. Grigorieva
preprint astro-ph/0210095.

\end{thebibliography}
\end{document}